\documentclass[letterpaper, 10 pt, conference]{ieeeconf} 
\usepackage{graphicx}
\graphicspath{ {./} }
\usepackage{url} 

\IEEEoverridecommandlockouts               
\title{\LARGE \bf
A low-cost, Arduino-like development kit for single-element ultrasound imaging*
}

\author{Luc Jonveaux$^{1}$ 
\thanks{*This work was not supported by any organization}
\thanks{$^{1}$Luc is just a independant maker, reachable at 
    {\tt\small kelu124 at gmail.com}%
}
}

\begin{document}

\maketitle
\thispagestyle{empty}
\pagestyle{empty}

\begin{abstract}

An open-source software ecosystem for ultrasound imaging is widely available to developers, however, limited resources can be found on the open-hardware side. The focus of this work was to develop an easy-to-use platform kit (hardware and software) for providing the community a complete experimental setup for ultrasound imaging at a low cost, without the need of specific equipment. The goal of this work resembles the needs of medical systems in the 80's where analog techniques using single-sensor devices were prominent.

To this end, two open-source, arduino-like modules have been developed for building a simple, yet complete, single-channel analog front-end system, where all the intermediary signals are readily accessible by the user. A single-channel architecture avoids the beamforming overhead, though it limits the quality of the captured image, and brings simplicity to the system. 

The modules were tested using re-purposed ultrasound mechanical probes, as well as non-medical transducers. Furthermore, different digital acquisition systems were utilized for providing the images of interest. The developed modules can also be used in Radio Frequency (RF) projects, non-destructive testing and control projects, as well as in low-cost medical imaging projects.

\end{abstract}

\section{Introduction}

\subsection{Approach}

\begin{figure}
\centering
\includegraphics[width=.98\linewidth]{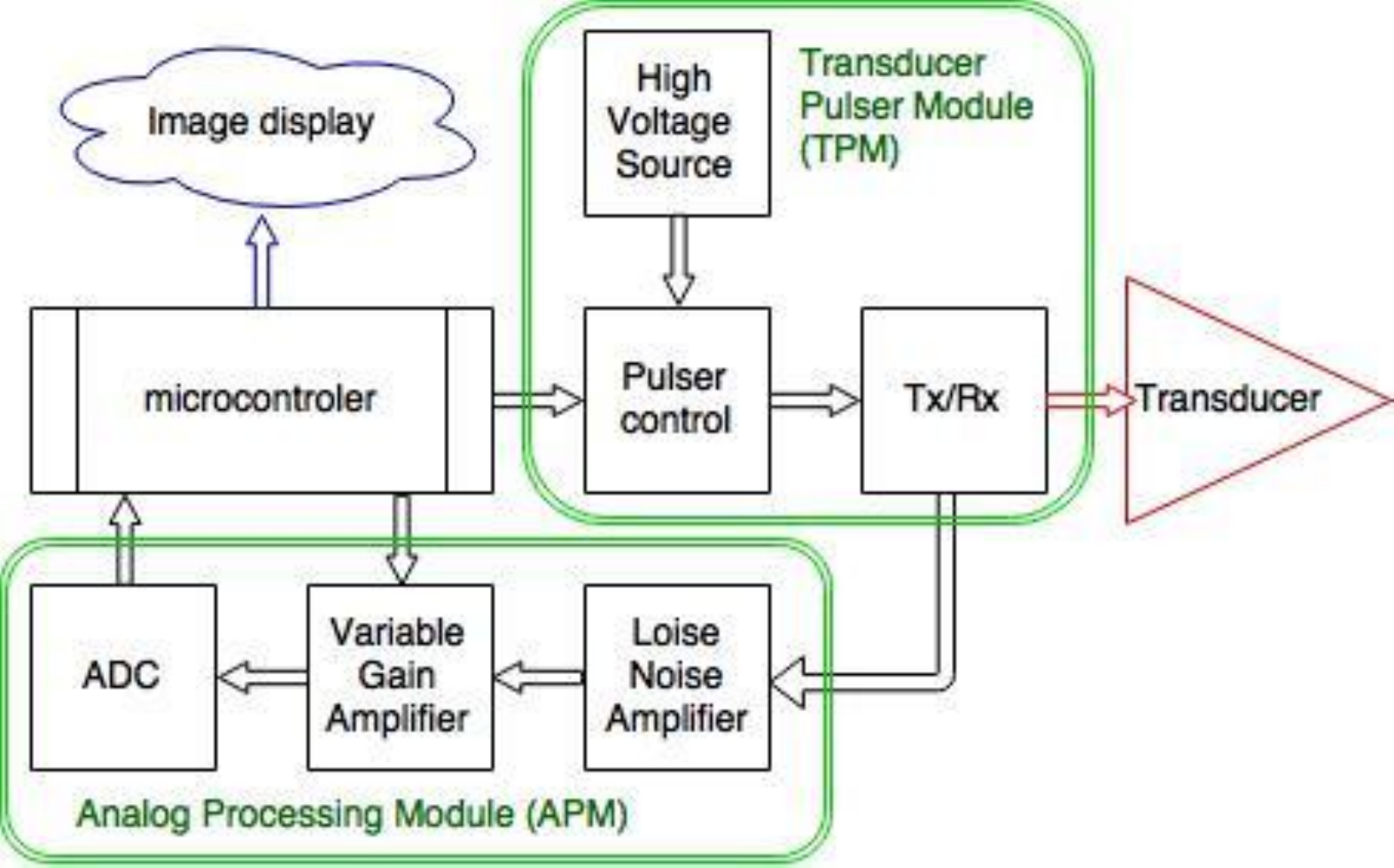}
\caption{Architecture of the ultrasound imaging system. The two custom boards, TPM and APM, are shown in green.}
\label{fig:UIAC}
\end{figure}

Ultrasound imaging has evolved since the first ultrasound machine appeared. The first devices were using single-sensor (transducers) techniques, coupled with mechanical scanning \cite{c25}. The architecture of such systems, as shown in Fig.~\ref{fig:UIAC}, is well-known and formed the basis of ultrasound imaging.

Mechanical scanning has its limitation, but also its strengths: a single signal channel, linked to a single sensor, means that the corresponding electronics are simplified, and the cost is reduced. Moreover, with progress made in different technical fields, mechanical probes are seen on the market again. Search in academic literature, and open-electronics resources, yielded little to no documentation of previous research to rebuild these mechanical ultrasound imaging devices.

To the best of the author's knowledge, there are no open-source hardware designs nor electronics accessible online for the analog-processing component. To bridge this gap, this work provides modules to the community to understand and recreate the electronic core of an ultrasound device. 

\subsection{This work}

This kit consists of several modules mainly built from easily available components. \textbf{Two electronic modules were specifically designed to provide the basic development kit}. These two modules, called the Ultrasound Imaging Analog Core (UIAC), as shown in Fig.~\ref{fig:UIAC}, are:

\begin{itemize}
\item the \textbf{Transducer Pulser Module (TPM)}: designed to provide a precise high-voltage pulse, necessary to excite the sensor, while remaining robust enough to be controlled by an Arduino;  
\item the \textbf{Analog Processing Module (APM)}: designed to correctly process the raw ultrasound electric signal, while easily exposing all intermediary signals, and exposing a digital output to the user. 
\end{itemize}

\subsection{A module approach}

We have chosen a modular approach to ensure that each key component inside the ultrasound image processing can easily be replaced and compared with another module. Each electronic module takes the place of a function in the signal processing chain or allows tapping into the different signals circulating between the blocks.

We have considered readily available open-source modules and recycled components (probes) to provide the user with building bricks for the basic non-medical ultrasound imaging tool. 

We remind that the aim of this work is not to design an ultrasound probe: \textit{the goal of the present article is rather to provide a basic open-source tool to understand ultrasound imaging technique and provide the analog core, unavailable today, as well as selecting proper off-the-shelf components for the other elements}.

This paper also discusses several possible options, keeping in mind that none are preferred and that the modular approach allows  different configurations. 

\subsection{Documentation}

A strong focus has been put on documentation and corresponding infrastructure. The project's documentation is backed by a script, extracting relevant information from the work logs in the repository, allowing a continuous update of information.

\section{Overall Implementation and design}

\subsection{Using echoes for imaging}

Ultrasounds, high-frequency sound waves, are used in medical applications for both diagnosis and treatment of patients. Their frequencies can vary from 2 to approximately 15 MHz for regular imaging, where in some cases higher frequencies are used for a finer surface imaging.

The ultrasound waves originate from the mechanical oscillations of a crystal in a transducer, excited by electrical pulses, also known as the piezoelectric effect. These pulses of sound are emitted from the transducer, propagate through the different media being imaged, and then return to the transducer as ``reflected echoes'' by an interface. These reflected echoes are converted back into electrical signals by the transducer and are further processed to form the final image.

\begin{figure}
\centering
\includegraphics[width=.8\linewidth]{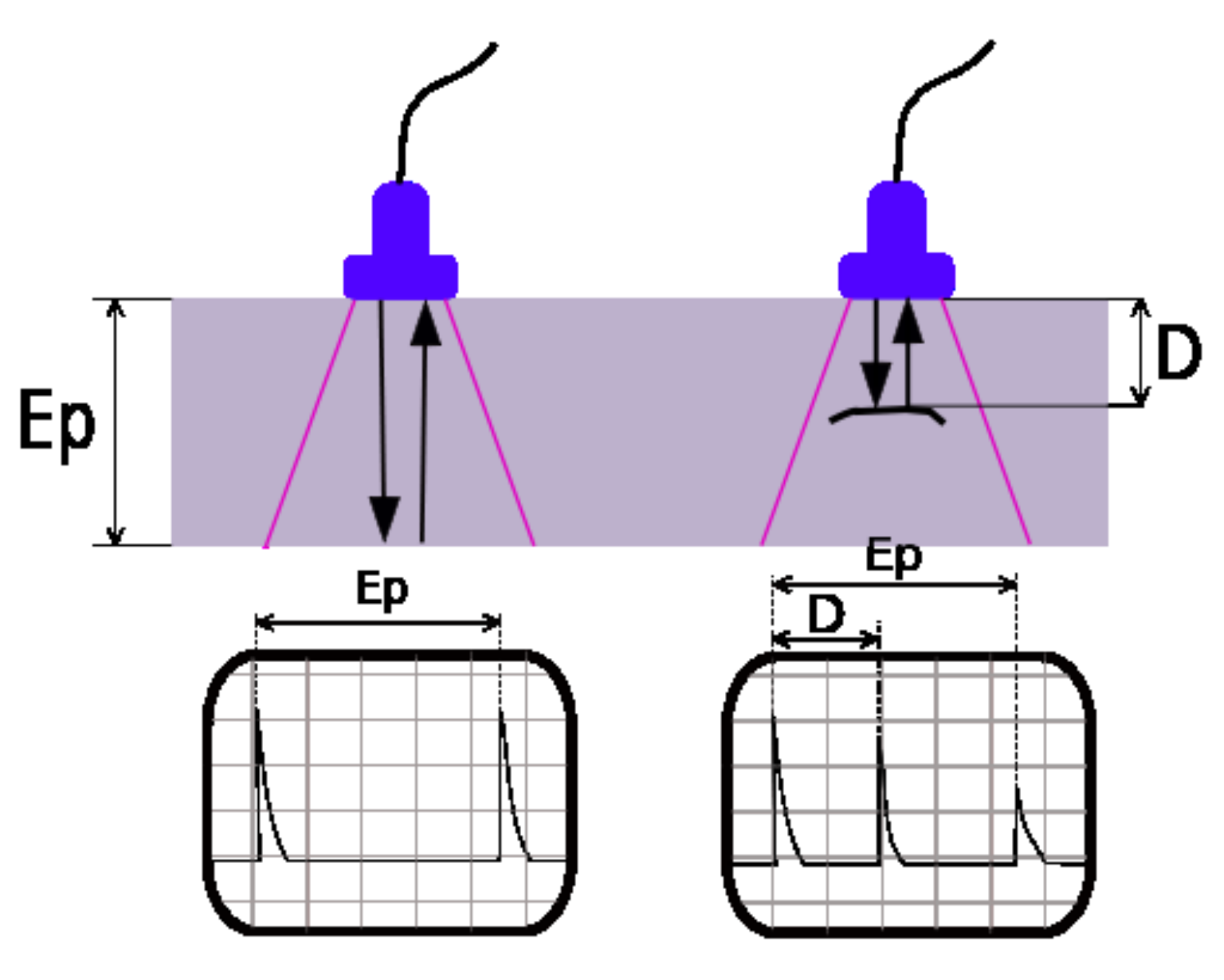}
\caption{Principle of ultrasound flaw detection. A void in a solid material reflects some energy back to the transducer, which is detected and displayed. Source: wikimedia.}
\end{figure}

In general, these sound waves, alike typical waves, are reflected at the interfaces between the tissues of different acoustic impedance (linked to the density of the medium), where the strength of the echo is proportional to the difference of the impedance. On the other hand, echoes are not produced if there is no acoustic difference, hence no impedance interface, between media. Homogeneous fluids are on the other hand seen as echo-free structures.

\subsection{A modularized approach}

With the aim of the kit being to allow one to explore the mechanisms of ultrasound processing, and to replace elements of the processing chain as desired, a modular approach was considered. Each module can be considered as a breakout board of the most central elements, intended for easy experimentation with usual equipment, such as breadboards and standard power supplies. This design, along with selected easily-accessed signal interfaces, provides access to the different intermediary signals.

\subsection{Requirements}

As mentioned, there is no available open-source electronic analog system for ultrasound imaging. There are several open-source software initiatives (like PLUS) or multiple-focus ultrasound control systems (like Vanderbilt's Open-Source Small-Animal MR-Guided Focused Ultrasound System), or articles suggesting electronic architectures for ultrasound systems. In 2009, Tortoli et al. [21] created an open 64-channel platform, but with a relatively complex architecture. A simpler 4-channel acquisition setup was built with an annular array, but with no automatic movement. A more recent approach combines relatively a 4-channel\cite{c22}, similar to our design, coupled to a Raspberry Pi. So far, state-of-the-art systems cannot be built with abundant modules and easily assembled components.

A representative publication shows a common structure for the different ultrasound systems [3], which has been summarized in Fig. \ref{fig:UIAC}. Recent progress and research show the feasibility of producing compact ultrasound imaging devices, which can interface with a smartphone \cite{c8}, wirelessly or through USB \cite{c20}.

A central element in the kit is the sensor, the ultrasound transducer. As we chose a single-channel processing unit, the transducer is a mono-element piezoelectric. This element has a fixed focal depth. For the sake of simplicity, a frequency on the lower range of ultrasound imaging was chosen. Larger characteristic periods enable simpler, slower controls and ADCs. A frequency of 3.5MHz was chosen, with a focal distance of 70mm, which is commonly used for obstetric and gynecological imaging. As a consequence, we chose a repetition period of 300us. This corresponds to an imaging depth of 230mm, more that was required to image between 20 and 150 mm. This project focused also on single element transducers to avoid developing a more complex beam-former component. The drawback is slow scanning, mechanical fragility, and insensitivity \cite{c12}. To increase the framerate, several transducers and corresponding connections can be integrated into a sweeping or rotating scan head \cite{c13}, as it was done on previous mechanical probes.

The kit requires a pulse component: to have the transducer emit a signal, a high voltage pulse, precisely controlled in amplitude and in time, needs to reach the transducer. We chose off-the-shelf components after bibliography and research. Existing ICs exist, such as the MAX14808 [8], but are relatively complex, as they are octal or quad chain components.

The kit then requires an ``analog processing'' component. After the acoustic wave leaves the transducer, echoes appear due to the acoustic impedance ruptures taking place in the medium being imaged. These echoes are captured by the transducer, and transformed back into a weak electrical signal, which needs to be processed. Classical processing includes filtering the signal around the central frequency of the transducer, then apply a low-noise amplification, then correcting the time-based attenuation. The image being the envelope of this last signal, one also needs to extract the envelope of this signal and pass this image to a digital converter. As filtering and band-pass consume most of the processing power (up to 80\% of processing power) \cite{c11}, the module allows signal envelope detection with an ADL5511 IC. This also enables a first compression of the data to be transmitted to the user.

On the digitization side, an echo being typically a couple of periods long, the envelope of the signal, hence the ultrasound image, would have a 1 MHz frequency, which would require specific ADCs. Open hardware boards have onboard ADCs, but very few have ADCs above the Msps range, This implies that an analog envelope detection has to take place in the module. However, we have used an existing open-source 40Msps Data Acquisition (DAQ) on one of the modules, and have as well tried a 6 Msps Arduino IDE-compatible micro-controller to acquire the signal and stream it over WiFi.

\subsection{A embedded-linux first iteration}

A first iteration of the hardware was embodied in a Beaglebone Black extension (cape), where the cape 10 Msps ADC would be tapping into the two programmable real-time unit (PRUs) to acquire the signal. Special attention was given to simplify power supply, limiting the inputs to 5V and 3.3V, the most common levels.

\subsection{Designs of modules}

Three iterations were done during the length of the project: 

\begin{itemize} 
\item A first single-board analog design was tested with a BitScope.
\item The second design used both the two TPM and APM modules, with a medical probe on one side, and a Beaglebone Digital Acquisition board on the other.
\item The third design is simpler, using the two modules, with a simple transducer for the sensor, and a Arduino-compatible microcontroller, with an appropriate ADC, and with wireless capabilities.
\end{itemize}

\subsection{A first integrated board}

This first iteration permitted tests and validated parts of the design. Despite its test points, this board did not provide all the insights that can be extracted from the hardware, so a redesign was considered to expose all key inputs and outputs of the signal processing. While existing compact elements exists, such as the AFE5808 (which include low noise amplifiers (LNA) and analog-to-digital converters (ADCs) \cite{c8}, we preferred using separate ICs, so that the user can measure exactly the target signal.

\begin{figure}
\centering
\includegraphics[width=.95\linewidth]{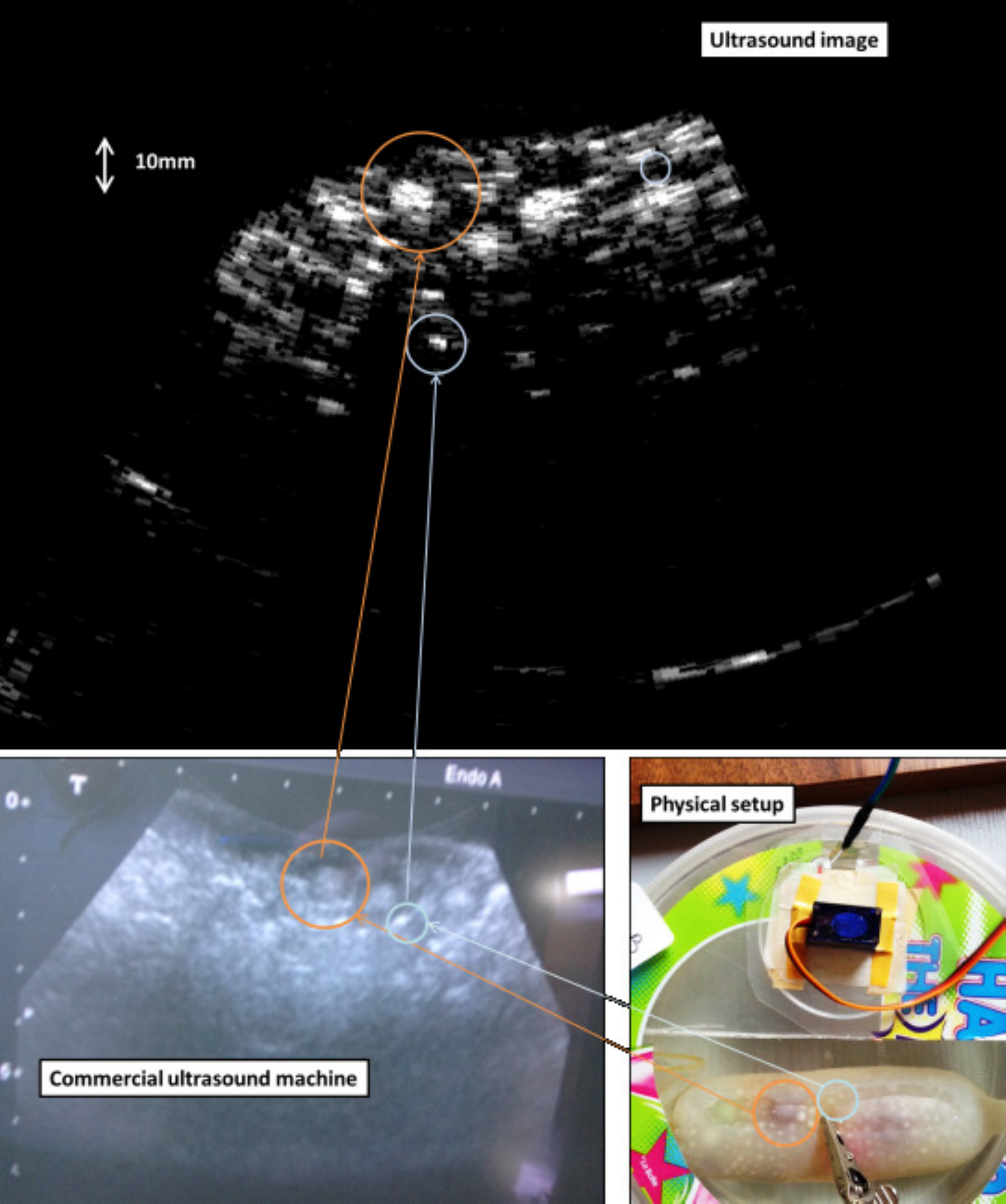}
\caption{Image obtained with the first iteration, on a Beaglebone cape design, data being captured with a Bitscope Micro at 5 Msps (setup shown in Fig~\ref{fig:workspace}).}
\label{fig:single}
\end{figure}

\subsection{A second approach}

For the sake of simplicity, a design of two separated modules emerged.

\begin{itemize} 
\item One is the Transducer Pulser Module (TPM), where the high voltage and connection to the transducer lies.
\item The other is the Analog Processing Module (APM). A dual input for the clipped raw signal from the transducer was integrated, as well as different jumpers and pots, to control the Variable Gain Amplifiers (VGA) gain, as well as the ADC reference voltage. The high-speed ADC was removed, and replaced with an on-board serial 2Msps ADC.
\end{itemize}

The remaining elements of the kits are micro-controllers, processing units, or power supplies that can be easily obtained and easily modified and programmed.

To test the two APM and TPM modules, we used a commercial ultrasound probe and a Beaglebone-based acquisition board.

\subsection{A third iteration}

The latest set of modules is based on a wireless-enabled, Arduino-compatible STM32, which has a 6 Msps on-board ADC, strictly compatible with the analog processing module for the envelope detection.

\subsection{A selection of integrated ICs}

To save on costs and complexity, and to ensure the robustness of the designs, the two designed modules leverage existing ICs:

\begin{itemize}
\item The Transducer Pulser Module (TPM) requires both a high-voltage source, and a pulser control. These functions used the Recom Power R05-100B DC/DC Regulated Converter with single-output, to generate a stable high-voltage, which level is determined by a potentiometer, and a Supertex HV7360, High Speed Two or Three Level Ultrasound Pulser, to precisely control the pulse level and duration.
\item The mono-channel Analog Processing Module (APM) uses a single channel ultrasound Time Gain Compensation (TGC) integrated circuit, the Analog Devices AD8331, Ultralow Noise VGA with Preamplifier, which gain can be controlled by an external 0 to 1V track. The amplified signal is fed into a RF envelope detector, the Analog Devices ADL5511, RF envelope and TruPwr™ rms detector. The envelope is the unbiased with an Analog Devices AD8691 Series Precision Amplifier, and optimized for the last item, the  Analog Devices AD7274, a 12-bit, 3 Msps Analog to Digital Converters.
\end{itemize}

\section{Quality control}

\subsection{Safety}
Most of the modules are found on usual open-source hardware procurement websites. However, the two modules specifically developed for this project fail to enter this category.

One of the modules uses a DC-booster, which can raise the voltage it delivers above 100V, as the R05-100B can deliver up to 134VDC. The design has limited high-voltage to specific points within the module, and to the SMA connector going to the transducer. The other pins of the module have inputs/outputs that range in the [-5V ; 5V] bracket. The full setup, without the motor, did not use more than 170 mA at 9V during the tests. The ATL3 probe, powered at 3.3V, brought the total power envelope to 330mA at 9V. The difference of stimulating the transducer caused a 5 mA at 9V difference.

Other safety issues when applied to humans \cite{c24}. There are indeed very strict regulations on ultrasound equipment for use in humans due to both thermal and mechanical effects of ultrasound on tissue such as overheating, and mechanical breaking of tissue structures. We understand that no medical scanners can be sold on the market without having undergone strict measurements and reporting routines (e.g. the FDA in USA and similar institutions elsewhere). 

However, the present modules do not aim at providing a medical ultrasound imaging system to a doctor, nor is suitable for tests on living beings, and hence does not need to fulfill the same requirements as a medical device.

\section{Calibration}

\subsection{Emulating the signal}

In order to obtain repeatable signals, the DAC of a Arduino-compatible board was used and integrated on a separate module. This module is capable of simulating a 2MHz signal, using an arbitrary signal profile. This signal can be used to characterize the analog processing module, as well as the DAQ modules. 

\subsection{Calibrating the signal}

Calibration of ultrasound electrical signal processing requires a standard signal, which is difficult to provide using a classical transducer. Moreover, due to the variety of transducers, it would be extremely difficult to obtain a standard setup and signal. A calibration tool has been built (the DAC module), based on a STM32F205 DAC. This permitted to record the calibration signal, and to simulate the behavior of the Transducer Pulser Module (TPM).  We note that do-it-yourself signal ultrasound generators \cite{c23} exist, though these were too complex for our purposes.

Qualification and calibration of the high-voltage level on the TPM are done using an oscilloscope.

Finally, the gain level of the analog processing module is in the tests described below adjusted once and for all at the beginning of the experiment, to use the full range of output signal and obtain the best image possible. The level of gain can be selected, either from one input signal, coming from an external DAC for example, or with a potentiometer, the choice between the two being made possible by a jumper.

\subsection{Developing a home-made reference material}

\begin{figure}
\centering
\includegraphics[width=.9\linewidth]{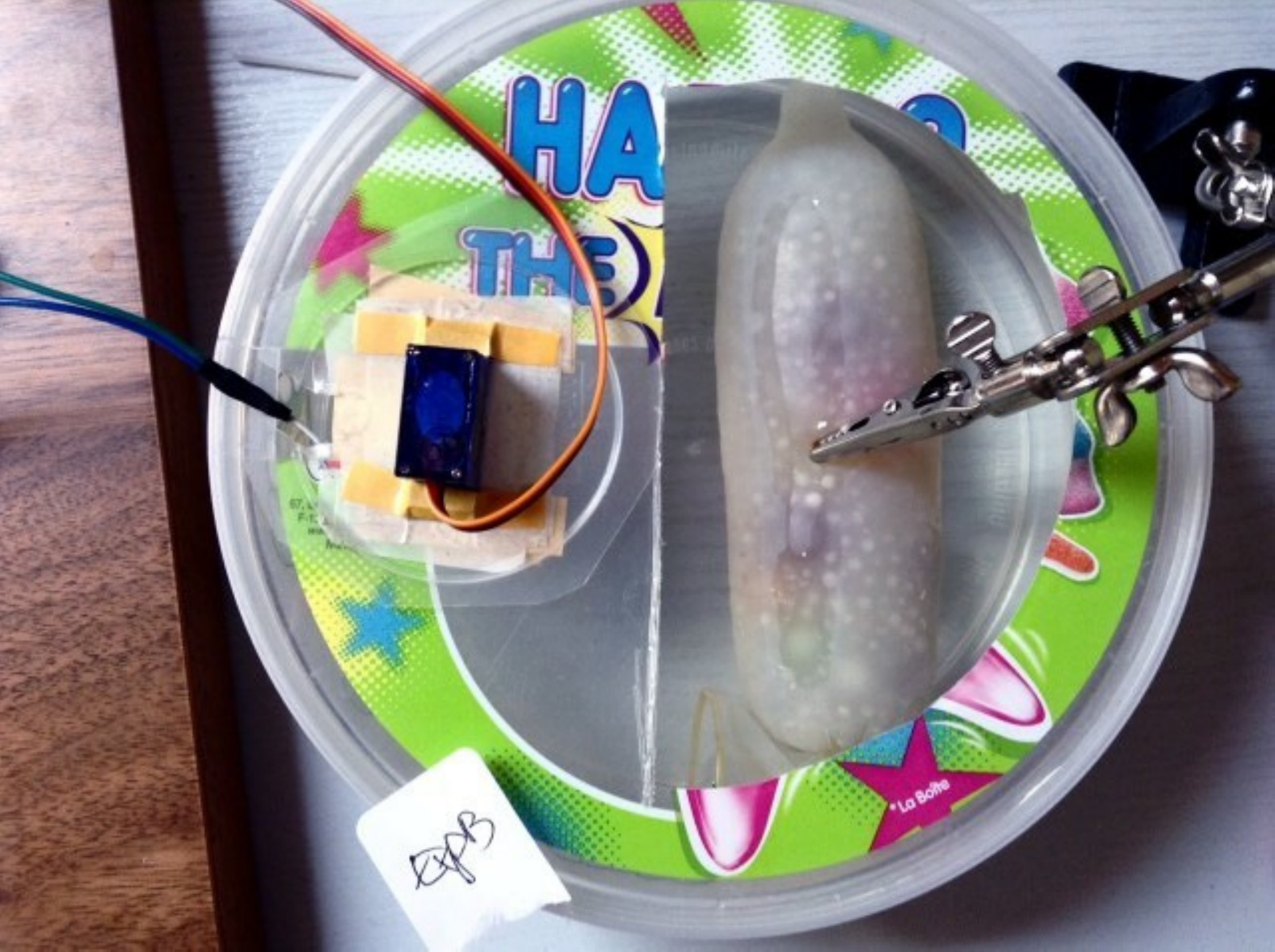}
\caption{The custom-made phantom.}
\label{fig:phantom}
\end{figure}

A home-made phantom was used to test the first iteration of the board. It was made of a gelatin phantom, with tapioca inclusions of two types (2mm and 8mm), the medium being contained in a condom (see Fig~\ref{fig:phantom}). This type of phantom does not conserve well, and was not reused for the second iteration. Tapioca and gelatin are used as they are close in terms of ultrasound properties to a human body.

\subsection{Developing a wire phantom}

\begin{figure}
\centering
\includegraphics[width=.8\linewidth]{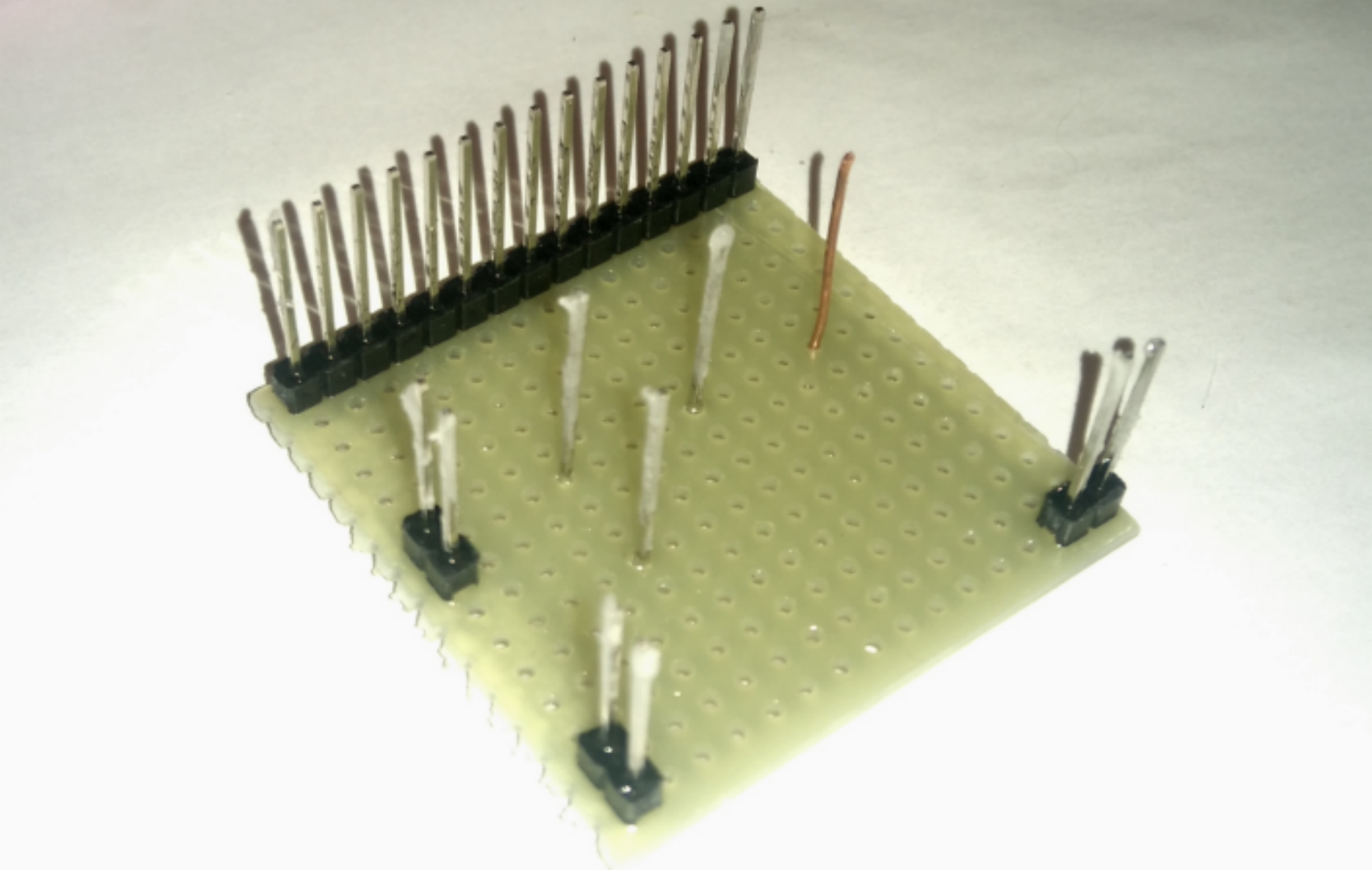}
\caption{The wire phantom used to determine the resolution of the device.}
\label{fig:wirephantom}
\end{figure}

The gelatin phantom is useful in terms of obtaining volumetric images. However, this type of phantom does not conserve well, and does not provide repeatable images. To this purpose, we have developed a very simple strip-board phantom, made of off-the-shelf electronic components, see~Fig.~\ref{fig:wirephantom}. 

\subsection{General testing}

General testing has been done, especially for the two modules:

\begin{itemize}
\item For the Transducer Pulser Module (TPM), the criteria of the tests were the duration of the pulse, which should match the input signals, as well as the voltage of the pulse, being set by a potentiometer. Tests were done using an oscilloscope.
\item For the analog processing unit, the DAC module allowed a standard input to be processed, and the result of the processing analyzed.
\end{itemize}

General conditions of the tests were that of a 150 ns, 70V-pulse for the Transducer Pulser Module (TPM), with a repetition every 300us. The gain setting on the analog processing board was set to maximize the range of the signal and match it to the DAQ unit.

\section{Application}

\subsection{Use case(s)}

These modules can be used in any setup requiring the excitation of the transducer, and further receiving and processing of the ultrasound signals. In our case, we have shown that they are working with commercial probes as well as single elements.

\subsection{General use case}

The modules can be assembled in a minimum set of additional modules using a power-supply, a pulser controller, and data acquisition.

\begin{figure}
\centering
\includegraphics[width=.95\linewidth]{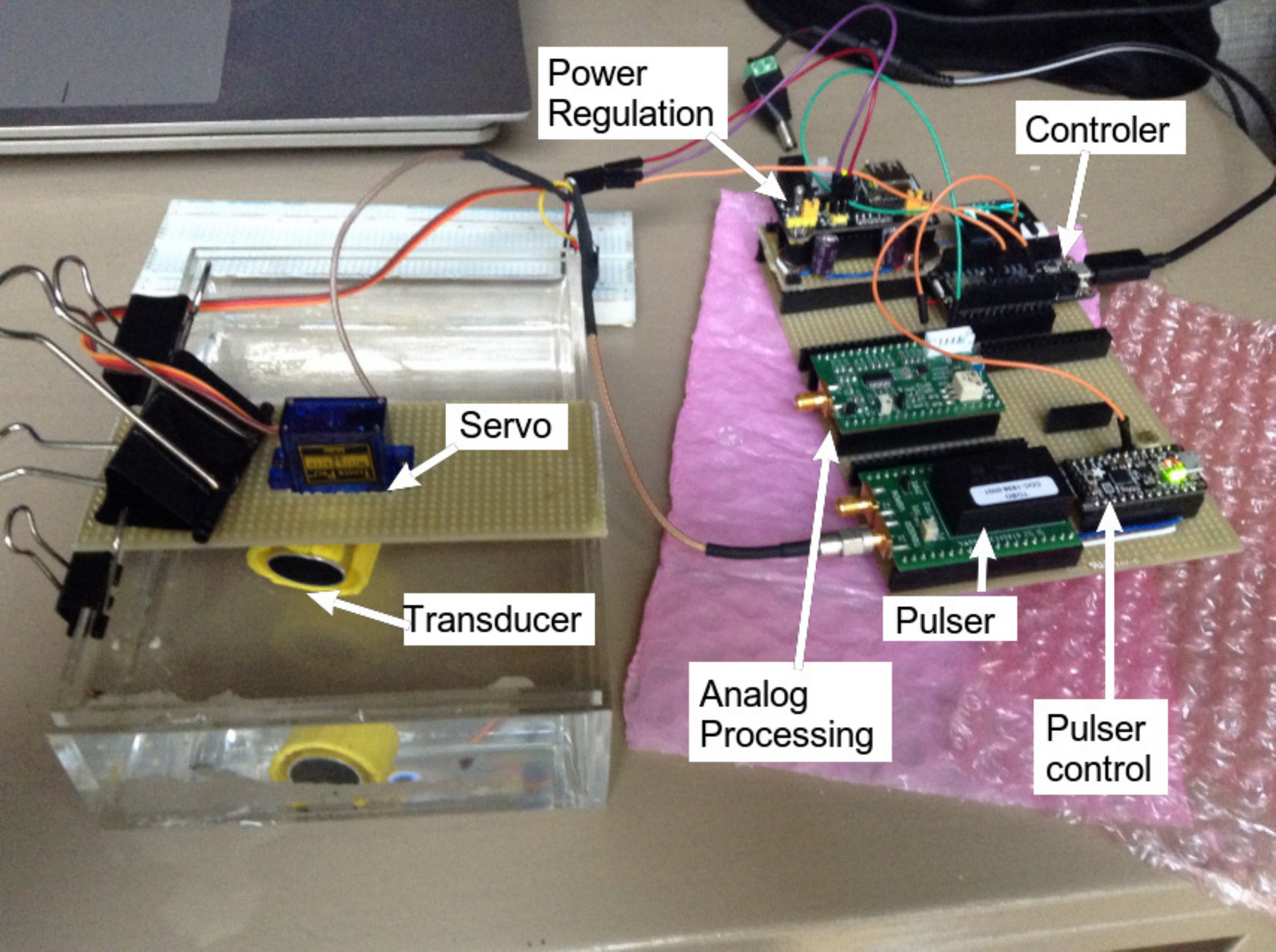}
\caption{Setup with a single element piezoelement, the two custom modules, and a Arduino-IDE-compatible micro-controller, streaming data over wifi.}
\label{fig:wireless}
\end{figure}

\begin{itemize}
\item The Beaglebone cape setup is relatively simple, see Fig~\ref{fig:workspace}. The pulse control was realized using a Trinket Pro (point A), controlling the board (B). The data was acquired with a Bitscope.

\begin{figure}
\centering
\includegraphics[width=.95\linewidth]{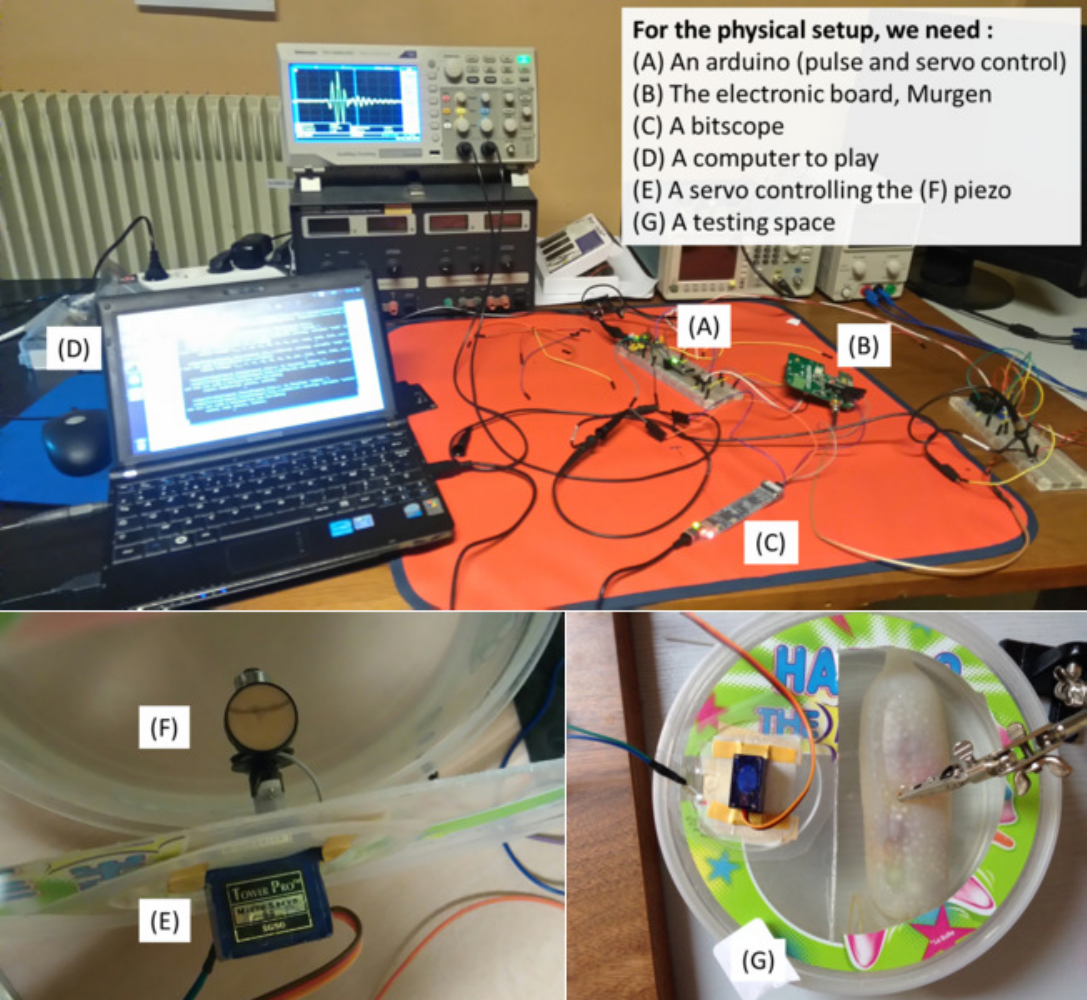}
\caption{The work space, with the first iteration, a salvaged piezoelectric element, and a servo motor imaging a custom-made phantom.}
\label{fig:workspace}
\end{figure}

\item For the second iteration iteration, the modules were used with a vintage probe found on eBay and the Beaglebone PRUDAQ cape, replacing the Bitscope, see Fig~\ref{fig:atl}.

\begin{figure}
\centering
\includegraphics[width=.95\linewidth]{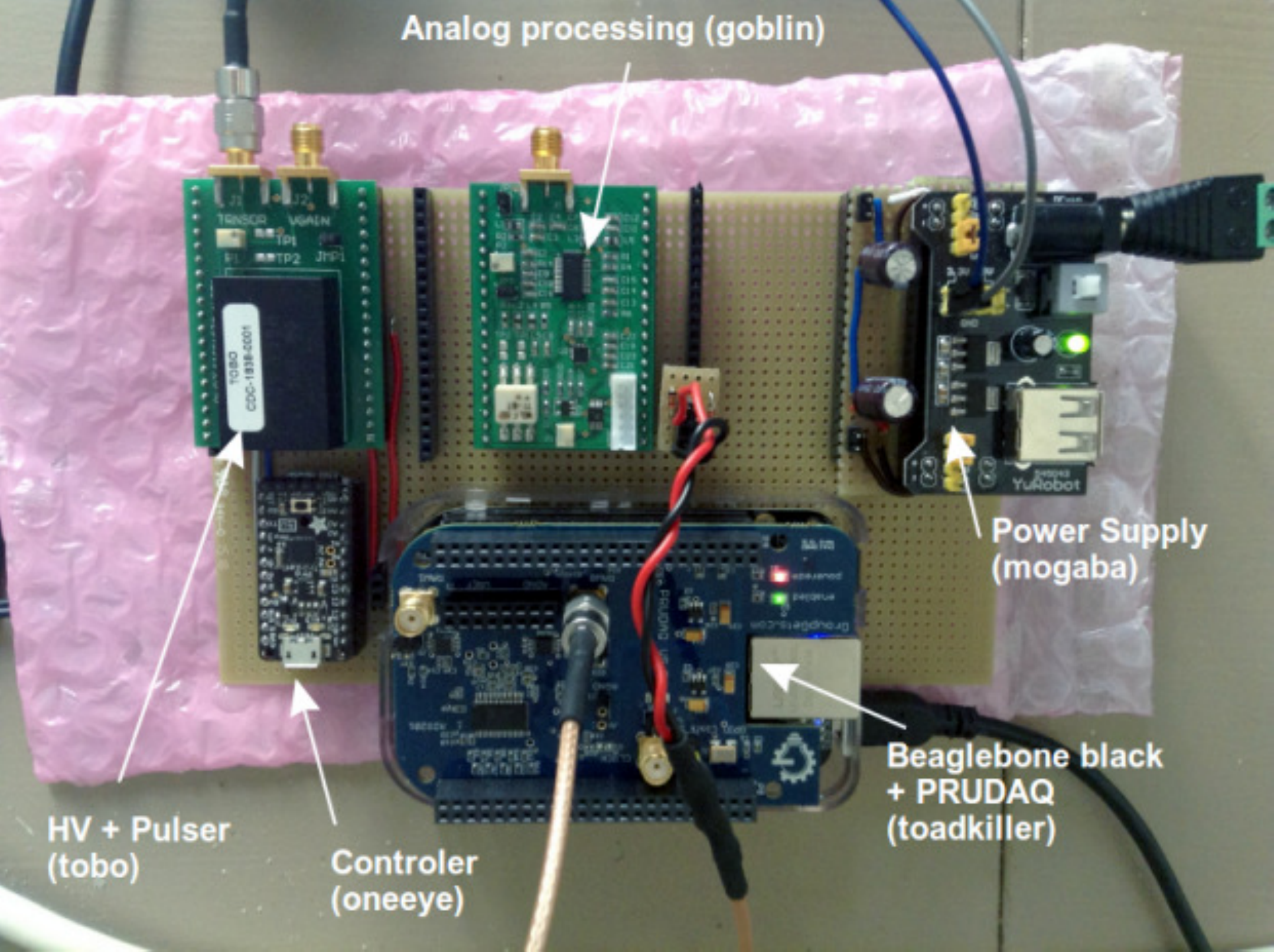}
\caption{The setup with a ATL probe, along with a PRUDAQ cape and the two home-made modules.}
\label{fig:atl}
\end{figure}

\item With the wireless setting, the setup simplifies, and only need to be powered with a USB cable (see Fig~\ref{fig:wireless}).
\end{itemize}

\subsubsection{Testing with single element transducer}

A first test was done using a single-element transducer, salvaged from a used endovaginal S-VRW77AK ultrasound probe. The transducer was moved by a basic servomotor controlled by a Trinket Pro 5V, which doubled as the controller sending the pulse commands. Data was acquired on the first iteration at 5Msps with a Bitscope unit (BS10). The noise kept the SNR low, but nonetheless provided a source of data, and permitted to assess the performances of the different blocks. See Fig.~\ref{fig:phantom} for the phantom, and Fig.~\ref{fig:workspace} for the setup. Image type is detailed in Fig.~\ref{fig:single}.

\subsubsection{Testing using a commercial transducer}

The dev-kit has been tested with a ATL-3 probe found online. This mechanical probe has 3 rotating transducers, a characteristic of this series of ATL probes. The ATL Access 3 probe connector to the transducers is simple and consists of a BNC cable. The other connector pins may correspond to the control of the probe motor. The probe's motor was connected to a 3.3V, allowing the transducers to rotate freely in the body of the probe. The probe also includes outputs exposing the motor control, ensuring that the excitation of transmitter pulses are synchronized with the rotation, so that each line appears in the same direction in space for each rotation of the probe. In our case, we have seen that the transducers rotate relatively uniformly with a 3.3V supply from the voltage regulator. Minor adjustments in images were done in software to obtain a consistent video loop.

\begin{figure}
\centering
\includegraphics[width=.8\linewidth]{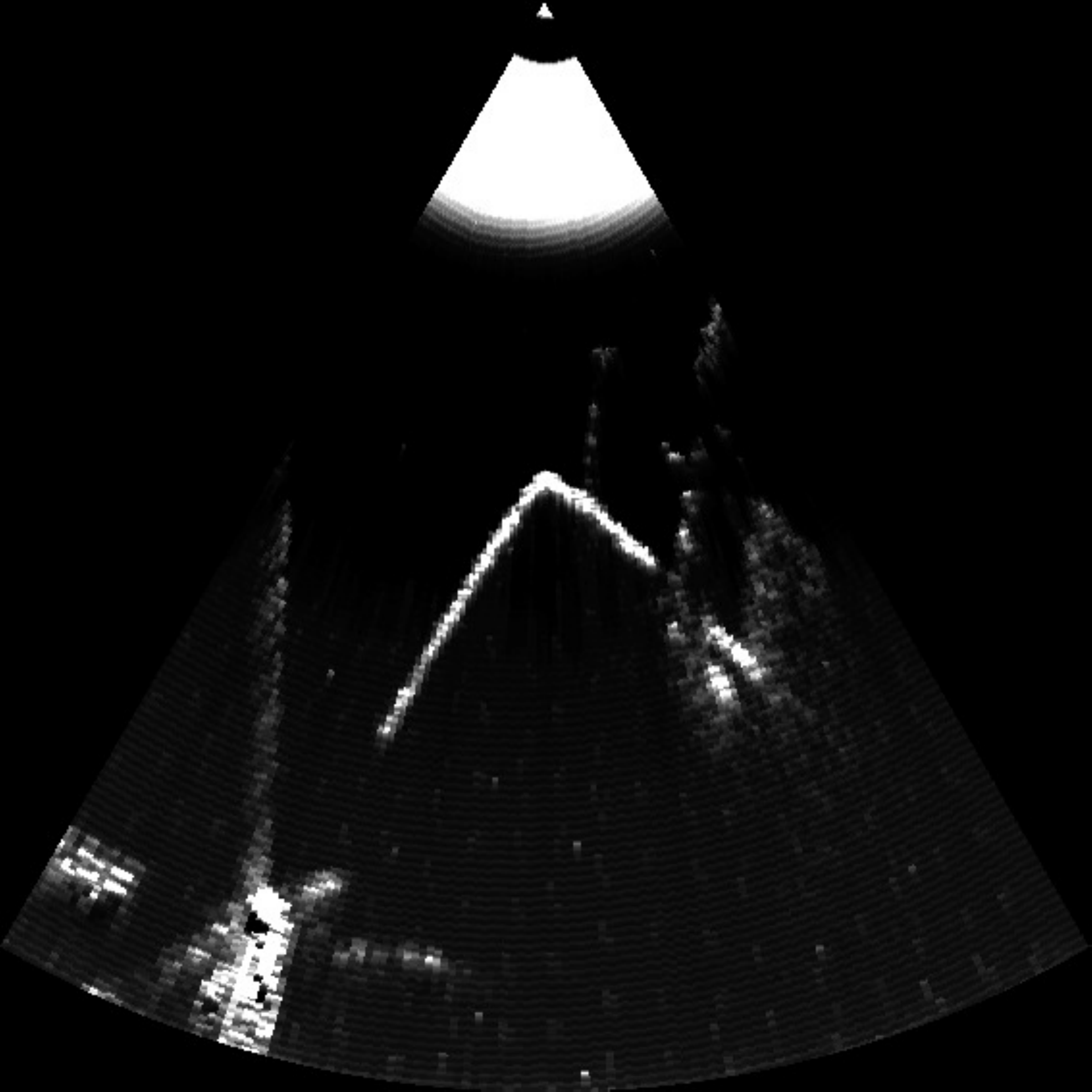}
\caption{Image sample acquired with the high-speed acquisition PRUDAQ and a ATL 3 probe (setup in Fig~\ref{fig:atl}).}
\label{fig:retroATL}
\end{figure}

A first test took place using a Bitscope micro (BS10), similarly to the previous application. However, the time interval between two captures was not consistent due to buffer transfer delays and did not lead to satisfactory acquisitions. A second try was done using the PRUDAQ, a faster, real-time acquisition unit, designed as an extension of the Beaglebone. The PRUDAQ was connected to the amplified envelope signal provided by the APM, clipped at 1.4V to protect the DAQ input. 32MB were acquired. With 32MB containing two signals encoded over 16 bits each, there are 8388608 points of data, or approximately 839ms. With a pulser set at 300us intervals, we should see around 2796 lines of data, which was the case. Rebuilding the image, we could see close to 12 full images, for a framerate of 14 fps, meaning that the motor was at approximately 290rpm. The setup was used on a ultrasound phantom, as well as on a small dice (10mm side), as seen in Fig.~\ref{fig:retroATL}.

\subsubsection{Testing with a Arduino}

\begin{figure}
\centering
\includegraphics[width=.98\linewidth]{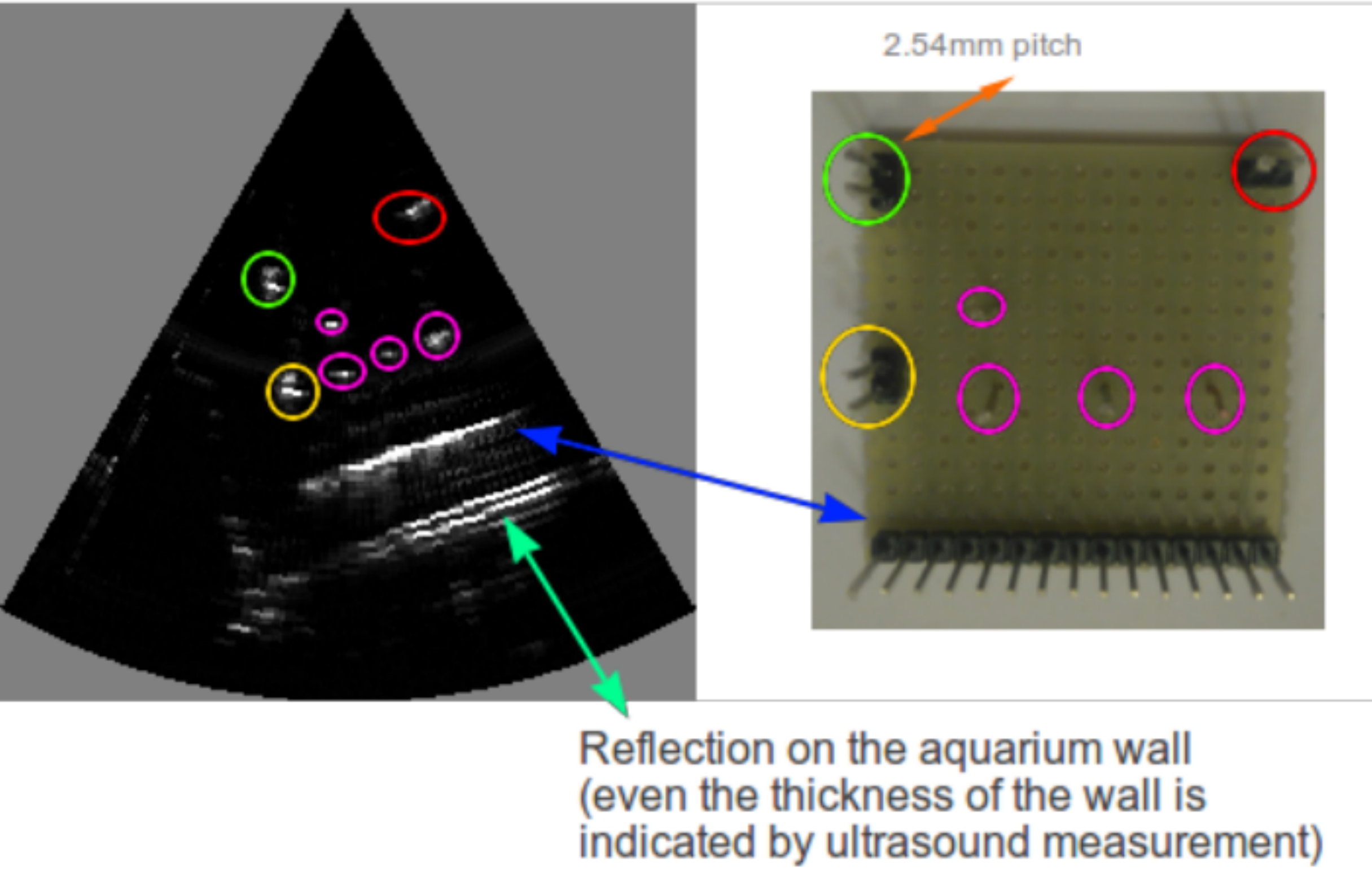}
\caption{Image sample acquired from the wire phantom with a simple transducer and the 2 Msps arduino-like, setup detailed in Fig~\ref{fig:wireless}}
\label{fig:wirelessimage}
\end{figure}

The last test was done using a piezo-element controlled by the Adafruit Feather WICED. The image obtained is shown in Fig.~\ref{fig:wirelessimage}. We show that with a 2 Msps acquisition, we can easily resolve the pitch between the different wires of the phantom. The full 6 Msps acquisition of the microcontroler still needs to be used.

\subsubsection{Datasets}

A small library of images obtained on home-made phantoms with the first iteration is available at \url{https://github.com/kelu124/murgen-dev-kit/tree/master/software/examples}.

Full data from the regular phantom imaging can be found on \url{https://github.com/kelu124/echomods/blob/master/include/20160822/2016-08-22-Fantom.md}.

The data corresponding to the image of the dice can be found at \url{https://github.com/kelu124/echomods/blob/master/include/20160814/2016-08-14-HackingAUltrasoundProbe.md}.

\section{Reuse potential and adaptability} 

\subsection{Reuse potential}

\subsubsection{Application for non-destructive testing (NDT)} 

The technology developed here does not differ from the technology used in NDT. Therefore, the whole set of modules can be used as-is in NDT, or its design can be adapted to NDT requirements.

\subsubsection{Application for transducer professionals} 

Often, ultrasound equipment repair professionals need to check each element in probe-arrays, and thus only need a single channel equipment.  

\subsubsection{High speed DAC}

Working with the DAC module enables one to be able to work on the acquisition system, based on a known signal input. This enables repeatability of the input, contrarily to images that are usually captured using a probe on an ultrasound phantom. This DAC can go up to 2MHz, on a limited voltage range, and could be used in testing analog signals processing. Recent development in open hardware provide interesting opportunities  \cite{c23}.

\subsubsection{Pulser}

It is suggested that the pulser can be used in medical ultrasound image devices, or for test applications. It's up to 35 MHz operating frequency and 2 ns matched delay times allow higher frequencies uses, such as superficial imaging or doppler analyses \cite{c1}.

\subsubsection{Analog processing}

The analog processing module was tested for signals from 2MHz to 10MHz, with some distortion happening on the higher end of the bracket. However, it can be noted that, with the possibility to measure and control the inputs and outputs between each processing unit, this module could possibly be used for signal demodulation.

\subsubsection{High-speed, wireless DAQ}

The wireless DAQ module we built can go up to 6 Msps, with a 10-bit resolution, on a [0;3.3V] input range, and can work on a battery, as there is a battery management system already built-in. This module could, therefore, be used in systems where rapid acquisition and wireless streaming need to be used. For example, for Software Defined Radio (SDR), this module can be used after demodulation.

\subsection{Adaptability of the home-made modules}

\subsubsection{Other uses}

The modules source code has been released, it is relatively easy for electronic engineers to reuse this code. 

Apart from the well-known obstetrics and gynecology uses, ultrasound devices have been developed for several uses, which the current hardware could support, such as: doppler ultrasonography, contrast ultrasonography, molecular ultrasonography, elastography, non destructing testing, bladder measurement and other. An interesting one, suggested by Richard et al \cite{c3}, would be to provide visual biofeedback to stroke victims relearning to control muscles. The same board could be reused for work on sonar-like systems \cite{c14}, or as an acquisition device for bone conduction microphones.

\subsubsection{A low-cost option}

Apart from the wireless-enable micro-controller at 35\$, the two custom modules components bill of materials cost 60\$ and 85\$ respectively. This low-cost lifts the barrier of high-cost equipment purchase and hence facilitates the reuse of these designs.

\subsection{Support} 

This project benefits from an infrastructure that is completely open: GitHub for the storage of files, and a GitBook to synthesize the complete documentation. A mailing-list gathers the users community, enabling a community-based support. Communication between the different projects is to be developed. 

\section{Build Details}

\subsection{Availability of materials and methods}

Most of the modules can be sourced from usual, well known open-hardware online suppliers. Moreover, the two specific modules can either be produced with a proper surface mount equipment, or the manufacturing files sent to a fab. The HV7360 becoming obsolete, a new version of the board has been published.

\subsection{ Ease of build}

When the first iteration had 4 layers, the two newly made modules are 2-layer designs, for the sake of simplicity.

The design also relies on off-the-shelf ICs to limit the number of its components. A trade-off had to be found when some ICs were BGA ICs.

The non-custom modules can be found commercially or already built.

Most of systems are FPGA (or DSP) based, especially for higher imaging frequencies \cite{c4,c5}, as well as using multiple-element transducers \cite{c7} -while maintaining costs and power consumption low. We considered that programming FPGA (even DSPs) was a steep requirement for non-specialists, hence we focused on alternatives. However, it can be noted that an FPGA module could be developed to interface with existing modules. 

\subsection{ Operating software and peripherals}

The hardware has modules that require software to operate. These modules rely on the Arduino IDE, and their code was compiled using Arduino IDE 1.6.9. For the wireless module, the WICED BSP version used was 0.5.5.

To collect the data, the Beaglebone module simply uses the Beaglebone black with its PRUDAQ cape installed, where the data being acquired is available on a device (/dev/beaglelogic). With the wireless module, any wifi-enabled device can acquire the UDP stream. 

\subsection{Dependencies}

\begin{itemize}

\item Most of the processing code is using Python 2.7, which is GPL-compatible.
\item The Beaglebone module is using a Beaglebone Black, which is under a Creative Commons Attribution-Share Alike 3.0 license.
\item The Feather WICED module is Open Hardware, and Open Source for its software.
\item The code for the Arduino-compatible modules is developed under Arduino IDE.
\item The two boards developed under this project are following the Open Hardware TAPR license. 
\item The source documents for these two boards was originally developed using Altium (proprietary), but the source has been ported to KiCad, which is under a GNU General Public License(GPL) version 3.
\end{itemize}

\subsection{Hardware documentation and files location} 

The module approach that has been followed enables the posting of all code, source files, images, and general documentation in a single GitHub repository.

For each module, a Readme file presents the module, provides clear images of the module, the inputs and output of each module, and describes what is required to build and run the experimental setup.

\begin{itemize}
\item Name: Github repository for the ultrasound Arduino-like modules
\item Persistent identifier: \url{https://github.com/kelu124/echomods/}
\item Release: \url{https://github.com/kelu124/echomods/releases/tag/v1.0} 
\item Filetypes: both boards are available in Altium and in Kicad format
\item Licence: TAPR Open hardware license under which the documentation and files are licensed
\item Publisher: Luc JONVEAUX
\item Date published: 31/10/16
\end{itemize}

The full documentation, available as a GitBook, provides more details on the rationale of the designs, as well as more general comments on the setup and the author's worklog. It can be found at: \url{https://kelu124.gitbooks.io/echomods/content/}

\section{Discussion}

\subsection{Conclusions}

In this work, we presented the development of a cheap (\$400) set of modules for ultrasound imaging, leveraging on existing open-source hardware and integrated circuits. Their power and voltage requirements, fitting within a USB power envelope, can easily be supplied by off-the-shelf 5V power banks, and their small and lightweight design (A5 format) allows a great diversity in manipulation and configuration. In terms of image quality, the modules provide surprisingly high quality considering the level of complexity. The proposed modular design, as building blocks, will enable users to use the existing sets of modules or tailor them, if necessary, according to their own requirements.

\subsection{Future Work}

This set of modules shows that ultrasound imaging can profit from a usable dev-kit. Several points in this work, however, have to be reviewed, if not improved. Indeed:

\begin{itemize}
\item In general, the design of the boards can be greatly improved. For example, having only two layers on the current design may be a source of noise. Moreover, applying a RF net to the board or using a RF shield for the sensitive parts may be an idea.
\item The pulser-module design uses only two inputs and one high voltage source. However, the chip enables more complex uses as a pulser, which can be further explored.
\item The PRUDAQ has a real-time access to the digital information, as well as a Linux userspace. Further programming would enable the Beaglebone platform to be a real server and controller of the setup.
\item The modules are slightly too wide for a breadboard: an effort could be done to make the pins available on a standard breadboard.
\item A whole field left unexplored so far is that of the transducer. As the key sensor in the kit, it would be interesting to explore relevant technologies to develop a low-cost, good-enough transducer.
\item The transducer at the moment lies in water. For ease of working, probes or scan head will have to be developed. Further work is required to determine the acoustic window material and its thickness. Several works already give pointers in that sense \cite{c1}.
\item A multiplexer module can be used, to interface this single channel kit with an array probe. Doing this would permit to do synthetic aperture imaging, and to characterize as well each element in the array.
\item From a software point of view, the modules could be wirelessly controlled, leveraging the existing wireless communication channel, so that researchers can use a single unit for a laboratory, controlled from personal computers.
\item For these high frequencies, a robust scan head can be used to obtain 130 fps, with light-weight transducer, and magnetic drive mechanism [2]. A bimorph actuator would be sufficient to drive the imaging transducer \cite{c4,c5} immersed in the probe. The advantage of such a scanning device would be to precisely know the position of the line being imaged, while being cheaper, and more robust as there are no mechanical parts. Another alternative would be an ultrasound motor \cite{c6}.
\item An additional module could link the ADC with a USB interface, providing as well the power for the other modules. We have shown that our modules can be run with a power bank, and previous work show is it possible to stay within the USB power envelope [3].
\end{itemize}

Having a set of inexpensive Arduino-like modules will support the development of ultrasound imaging research, and provide the keys to the researchers, makers and curious-minded persons to explore this field.

\section*{Others}

\subsection*{Acknowledgements}

Huge thanks to the friends in the community for their sharing their ideas and giving their support. Thanks as well to Prof Charles and Zach to have given a try at testing the first iteration, Sofian for his help on the hardware, the Hackaday community for giving me the chance to go to the 2016 finals, Nicolas for his interest, the echOpen community (Benoit, Farad, Vincent, Jerome, Virginie, Emilie and the others) who have all kept me motivated!

\subsection*{Funding statement}

This project has been funded by personal funds, and supported by two prizes from the Hackaday 2016 contest.

\subsection*{Competing interests}

The author declares that he has no competing interests. Though LJ is a founder of the echOpen's project, this work has been pursued individually, the echOpen association having no involvement with this work.

\subsection*{Upcoming publication}

This document is pre-print version of a publication submitted for the Journal of Open Hardware (JOH).

\addtolength{\textheight}{-11cm}  



\end{document}